\documentclass[np2,twoside,twocolumn,showkeys]{revtex4}
\usepackage{bm,multirow}
\usepackage{kotex}
\usepackage{amssymb}
\usepackage{amsmath}
\usepackage{graphicx}
\usepackage{epsfig}
\usepackage{color}
\usepackage{bm,multirow}
\usepackage{hyperref}
\usepackage{ulem}
\usepackage{siunitx}
\usepackage{verbatim}
\usepackage{makecell}

\begin{document}

\title[Semantic Network Analysis of Physics Achievement Standards]
{Semantic Network Analysis of Achievement Standards\\ in Physics of 2022 Revised Curriculum}

\author{Jibeom Seo}
\thanks[E-mail: ]{qja1264@gmail.com}
\affiliation{Department of Physics, Sungkyunkwan University, Suwon 16419, Korea}

\author{Jun Cho}
\affiliation{Singil Middle School, Ansan 15401, Korea}

\author{Sukyung Han}
\affiliation{Namhan High School, Hanam 12978, Korea}

\author{Meesoon Ha}
\thanks[Correspondence to: ]{msha@chosun.ac.kr}
\affiliation{Department of Physics Education, Chosun University, Gwangju 61452, Korea}

\date[]{}

\begin{abstract}
We investigate semantic networks of achievement standards for physics subjects in the 2022 revised curriculum to derive information embedded in the curriculum. We extract each subject’s keywords with node strength and random-walk betweenness, detect communities of physics terms by the optimized greedy algorithm, and find the connectivity of physics subjects using bipartite networks. The network analysis reveals three remarkable results: First, keywords are about scientific thinking and practices, evolving to a higher level as the grades increase. Second, there is a lack of connection to learning content in physics. Lastly, achievement standards for ‘Integrated Science’ are inadequate to fulfill the intended purpose of the curriculum. This is attributed to the reduced learning volume in the 2022 revised curriculum. Our study implies that the curriculum and achievement standards should be improved for the better connectivity of subjects.
\end{abstract}

\keywords{2022 revised curriculum, Semantic network, Physics term, Achievement standard, Connectivity}

\maketitle
 
\phantom{for layout}
\vskip 20mm

  
\twocolumngrid
\section{Introduction}
The 2022 Revised National Curriculum in South Korea aims to achieve the educational goals of elementary and secondary schools. It provides guidelines for organizing and operating school curricula, as well as for teaching and learning methods\cite{2022}. 
A monitoring study analyzing the implementation of the 2015 Revised National Curriculum (the predecessor to the 2022 version) found that secondary school teachers perceived student-centered instruction and process-based assessment positively, and were attempting to apply them in their teaching. Both of these approaches are emphasized in the national curriculum\cite{2015}.
These findings highlight that the national curriculum significantly influences teachers' educational practices.

The national curriculum provides achievement standards for learning content. These standards represent the expected learning outcomes for students: what they should know and be able to do after mastering the essential content of a given subject. They serve as a guide for the learning objectives presented in textbooks\cite{L_K_2019}.
Moreover, because achievement standards outline the nature of the content, its connections to other topics, and the overarching direction of the national curriculum, analyzing them can provide a deeper understanding of a specific subject or learning content.
With the revision of the national curriculum, various studies have been conducted to analyze science achievement standards. For instance, some studies have compared the achievement standards of the 2015 and 2022 revised curricula in specific subjects\cite{K_2023, J_2023}, and one study has analyzed achievement standards related to the energy topic\cite{J_N_C_2023}.
However, previous studies have mainly focused on individual subjects or content areas, and research on the interconnections among subjects and topics remains limited. These connections are closely related to students’ academic achievement and interest\cite{Y_P_2014}. In particular, physics has a highly sequential structure, and a lack of prior knowledge can hinder learning\cite{L_L_2015}. Therefore, examining the coherence among topics within the field of physics may help address potential issues anticipated in the 2022 Revised National Curriculum.

One way to examine the connectivity among subjects within the field of physics is to analyze semantic networks. Semantic networks are powerful tools for identifying key terms in a text and understanding their meaningful interconnections. Furthermore, by analyzing their structure and quantitative properties, we can uncover new information not explicitly apparent in the original document.
In science education, previous studies using semantic network analysis have explored various topics.
These studies include examining the relationships between key scientific competencies in the curriculum and physics achievement standards\cite{R_N_I_2018}, and analyzing how achievement standards for specific content areas are described across different subjects\cite{J_N_C_2023}.
Other related studies have also been conducted\cite{J_2023, S_2018, S_K_2020}.

In this study, we aim to investigate the connectivity between subjects in the physics domain of the 2022 Revised National Curriculum. To achieve this, we will construct and analyze semantic networks using the words in each subject's achievement standards. Through this network analysis, we seek to derive various insights into the physics-related subjects.

\section{Research methods}
\subsection{Data source}

We analyzed the physics-related achievement standards within the 2022 Revised National Curriculum. For science subjects, the national curriculum is structured into two main phases. From elementary school (Grade 3) to middle school (Grade 9), it is implemented as a core curriculum under the subject \textbf{Science}. This foundational \textbf{Science} subject connects to content learned in Grades 1-2 and is designed to build essential scientific literacy for high school science studies. Within \textbf{Science}, the content domains of ‘Motion and Energy’ and ‘Matter’ are specifically related to physics.
For high school (Grade 10 to Grade 12), the curriculum transitions to a credit-based system tailored to students' career paths. This high school science curriculum comprises ‘core subjects,’ ‘basic elective subjects,’ and ‘career-focused elective subjects.’
\begin{itemize}
    \item Core subjects, including \textbf{Integrated Science 1 and 2} and \textbf{Science Inquiry Experiment 1 and 2}, are designed to foster scientific literacy, based on the foundational knowledge acquired in middle school.
    \item Basic elective subjects extend the ‘core subjects,’ focusing on foundational scientific concepts for students interested in science and engineering careers. The subject \textbf{Physics} falls into this category.
    \item Career-focused elective subjects are designed for students who specifically pursue science and engineering pathways. In physics, these include \textbf{Mechanics and Energy} and \textbf{Electromagnetism and Quantum}\cite{2022, I_2023}.
\end{itemize}

This study analyzed 127 physics-related achievement standards from the 2022 Revised National Curriculum. This included standards covering thermodynamics and statistical mechanics, as well as those from \textbf{Integrated Science 1 and 2} that relate to broader physics and science concepts. However, \textbf{Science Inquiry Experiment 1 and 2} were excluded since their standards do not explicitly describe physics content. Details on the selected achievement standards are as follows:
\begin{itemize}
\item For elementary school \textbf{Science}, 37 standards were selected: 20 standards from Grades 3–4 covering the topics ‘Forces in Everyday Life,’ ‘Objects and Materials,’ ‘Properties of Sound,’ ‘Uses of Magnets,’ ‘Changes in the State of Water,’ and ‘Various Gases’; and 17 standards from Grades 5–6 covering ‘Properties of Light,’ ‘Heat and Everyday Life,’ ‘Resources and Energy,’ ‘Motion of Objects,’ and ‘Uses of Electricity.’
\item For middle school \textbf{Science}, 26 achievement standards were selected, covering the topics ‘Heat,’ ‘Changes in the States of Matter,’ ‘Forces in Action,’ ‘Properties of Gases,’ ‘Light and Waves,’ ‘Electricity and Magnetism,’ and ‘Motion and Energy.’
\item For the core subjects in high school, a total of 13 standards were included from \textbf{Integrated Science 1 and 2}: 9 from \textbf{Integrated Science 1}, consisting of selected standards from ‘Foundations of Science,’ ‘Patterns in Matter,’ and ‘Systems and Interactions’; and 4 from \textbf{Integrated Science 2}, covering ‘Change and Diversity,’ and ‘Environment and Energy.’ In this study, \textbf{Integrated Science 1 and 2} were collectively referred to and analyzed as \textbf{Integrated Science}.
\item Finally, all achievement standards from the high school elective subjects \textbf{Physics}, \textbf{Mechanics and Energy}, and \textbf{Electromagnetism and Quantum} were included—18, 16, and 17 standards, respectively.
\end{itemize}

\subsection{Data analysis methods}
\subsubsection{Data preprocessing}

We extracted morphemes from each achievement standard, using the ‘large model’ of the Korean morphological analyzer Khaiii (Kakao Hangul Analyzer III)\cite{khaiii}. Among these morphemes (hereafter referred to as ‘words’), we only used general nouns (NNG) and proper nouns (NNP) for our analysis.
However, Khaiii often tokenizes words too finely, which frequently failed to capture the intended meaning of the achievement standards. To address this issue, we refined the extracted words based on a physics terminology glossary\cite{phyWord} and discussions with a co-researcher. We also removed meaningless or contextually irrelevant words and standardized synonymous expressions (e.g., ‘life’, ‘real life’, and ‘everyday life’ were all unified as ‘everyday life’).

\begin{figure*}[]
\includegraphics[width=\textwidth]{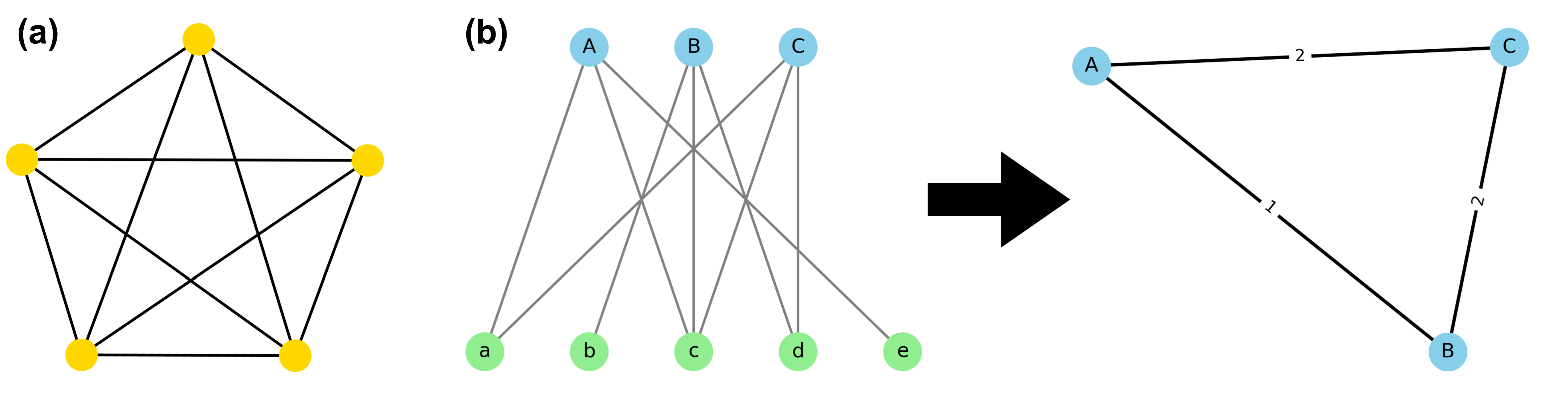}
\caption{(a) Clique with five nodes. (b) Bipartite network (left) and weighted one-mode projection of the network onto the upper node set (right).}
\label{fig:Bigraph}
\end{figure*}

\subsubsection{Semantic network}
A network, as shown in Fig.~\ref{fig:Bigraph}, consists of nodes and edges, where nodes represent points and edges denote connections between them. To construct a semantic network, we created a fully connected network, or clique, from each individual achievement standard, as illustrated in Fig.~\ref{fig:Bigraph}~(a). In these cliques, each unique word acts as a node, and an edge signifies a connection between two words. Duplicate words appearing within the same achievement standard were removed before constructing the clique.
As an example, consider the achievement standard: ‘We can predict the motion of an object by calculating the net force of various forces acting on the object.’ The extracted words are ‘predict,’ ‘motion,’ ‘object,’ ‘calculate,’ ‘net force,’ ‘force,’ ‘act,’ and ‘object.’ Even though ‘object’ appears twice, only its unique instance is used, resulting in seven words forming the clique.
Finally, we merged all these individual cliques to form the complete semantic network. We then categorized the words into \textit{all words} and \textit{physics-terms} and analyzed the differences between the resulting semantic networks.

To analyze the achievement standards, we used three key metrics: \textit{node degree}, \textit{edge weight}, and \textit{node strength}. Node degree measures the number of edges connected to a node, while edge weight indicates how many times two specific nodes are connected. Node strength is the sum of all edge weights connected to a given node.
For example, if the node ‘object’ connects to $x$ different words in the semantic network, its node degree is $x$. If ‘object’ and ‘motion’ are connected $y$ times, their edge weight is $y$. The node strength of ‘object’ is calculated as the sum of all edge weights $y_i$ between ‘object’ and the connected words $i$, i.e., $\sum_i y_i$. A higher node strength indicates greater importance within the semantic network.

We also used \textit{random-walk betweenness centrality} to quantify a node’s influence on information flow between other nodes. This value ranges between 0 and 1, with higher values indicating greater importance.
To illustrate, imagine information flowing from node $s$ to node $t$. If a node $i$ frequently appears along the various paths between $s$ and $t$, then node $i$ plays an important role in information transmission. This is the core idea behind betweenness centrality. While standard betweenness centrality assumes that information flows along the shortest path between two nodes, random-walk betweenness centrality assumes that information travels along random paths. Since the shortest-path assumption may not be appropriate for certain types of network structures\cite{RWB}, we adopted random-walk betweenness centrality.

In summary, node strength assesses how strongly a node connects to many others, while random-walk betweenness centrality measures its influence over information flow. By using these distinct metrics, we aimed to analyze the data from multiple perspectives.

\subsubsection{Community detection}
To detect communities formed by word clusters within the network, we utilized an \textit{optimized greedy algorithm} that seeks to maximize modularity\cite{OGA}. Modularity serves as a metric to assess the quality of a community division, defined as the difference between the number of edges within actual communities and the expected number of such edges in randomly generated communities. Since randomly generated networks are unlikely to exhibit meaningful community structures, higher modularity values indicate better community divisions. The maximum modularity value is 1.
Given that we are working with a weighted network where edges have weights, modularity is defined as follows\cite{1st}:
\begin{align}
Q_W = \frac{1}{W}\sum_C\left(W_C - \frac{s_C^2}{4W}\right),
\end{align}
where $W$ is the total sum of edge weights in the network, $W_C$ is the total sum of edge weights within a community $C$, and $s_C$ is the sum of node strengths within community $C$.
We used the \texttt{greedy\_modularity\_communities} function provided by NetworkX\cite{NetworkX}. Within this function, the \texttt{weight} parameter was assigned the edge weights of the network, while all other parameters were left at their default values.

\subsubsection{Bipartite network}
To analyze the connectivity between physics subjects, an undirected \textit{bipartite network} was used. A bipartite network consists of two distinct node sets, where connections are only allowed between nodes of different sets. 
Figure~\ref{fig:Bigraph}~(b) illustrates an example, showing an upper node set labeled with uppercase letters and a lower set with lowercase letters. If we imagine the upper nodes as subjects and the lower nodes as words, then word \texttt{a} appears in subjects \texttt{A} and \texttt{C}, while subject \texttt{B} contains words \texttt{b}, \texttt{c}, and \texttt{d}.
By examining the words shared between subjects, we can identify associations. For example, subject \texttt{A} and \texttt{B} share word \texttt{c}; subject \texttt{A} and \texttt{C} share words \texttt{a} and \texttt{c}; and subject \texttt{B} and \texttt{C} share words \texttt{c} and \texttt{d}. This information allows us to construct a \textit{one-mode projection} of the bipartite network onto the set of subjects, as shown in Fig.~\ref{fig:Bigraph}~(b).
In this one-mode projection, the relationships between subjects are quantified by the number of shared words. Consequently, subject pairs like \texttt{A} and \texttt{C}, and \texttt{B} and \texttt{C}, which share more words than \texttt{A} and \texttt{B}, are considered more closely related.

In particular, to examine the connectivity among subjects, we constructed a bipartite network in which the upper node set represents subjects and the lower node set represents words. From this, a projected network consisting only of subject nodes was generated.  
To represent the significance of connectivity between subjects, both \textit{edge weight} and \textit{random-walk betweenness centrality} for edges were used\cite{RWB, ERWB}.
While random-walk betweenness centrality is typically used to measure a node's influence on information flow, here, we adapted it to edges. This allowed us to compare and analyze the importance of the connections themselves within the projected network. A higher centrality value, which ranges between 0 and 1, indicates a more important corresponding edge.

\begin{figure*}[]
\includegraphics[width=\textwidth]{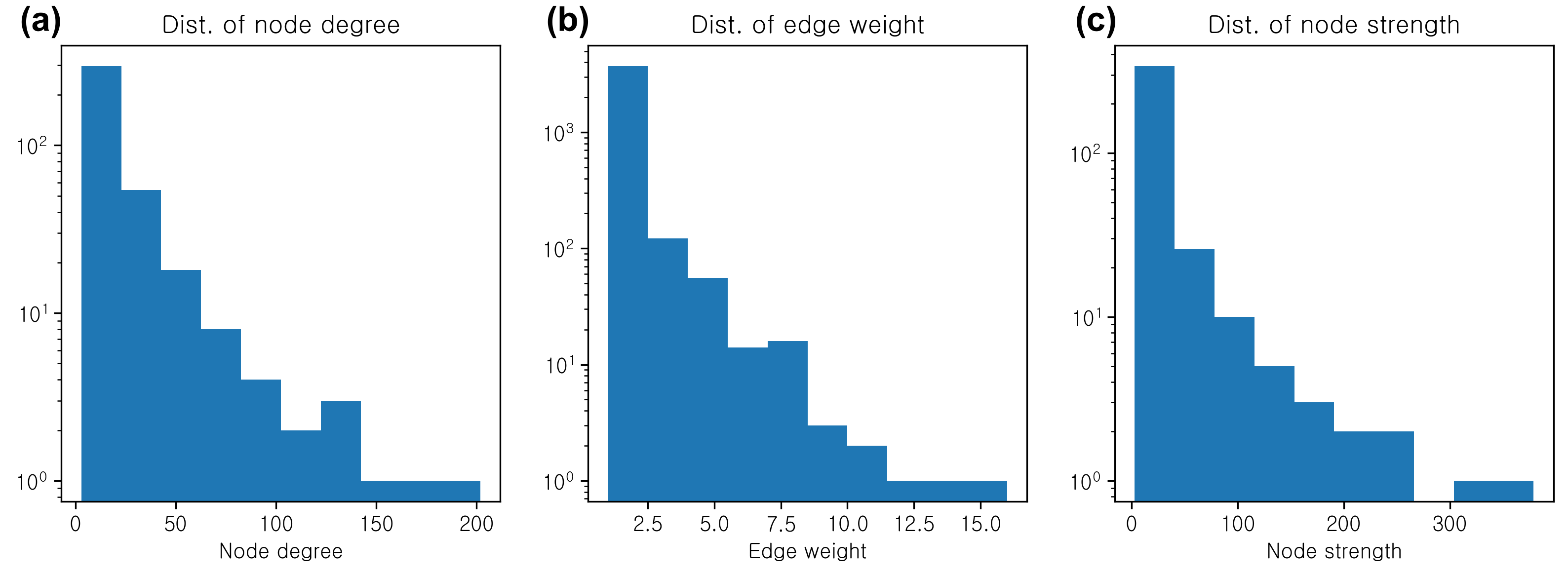}
\caption{For the whole word set from all physics subjects, the semi-log plots of (a) degree, (b) edge-weight, and (c) node-strength distributions.}
\label{fig:TotalStat}
\end{figure*}

\begin{figure*}[]
\includegraphics[width=\textwidth]{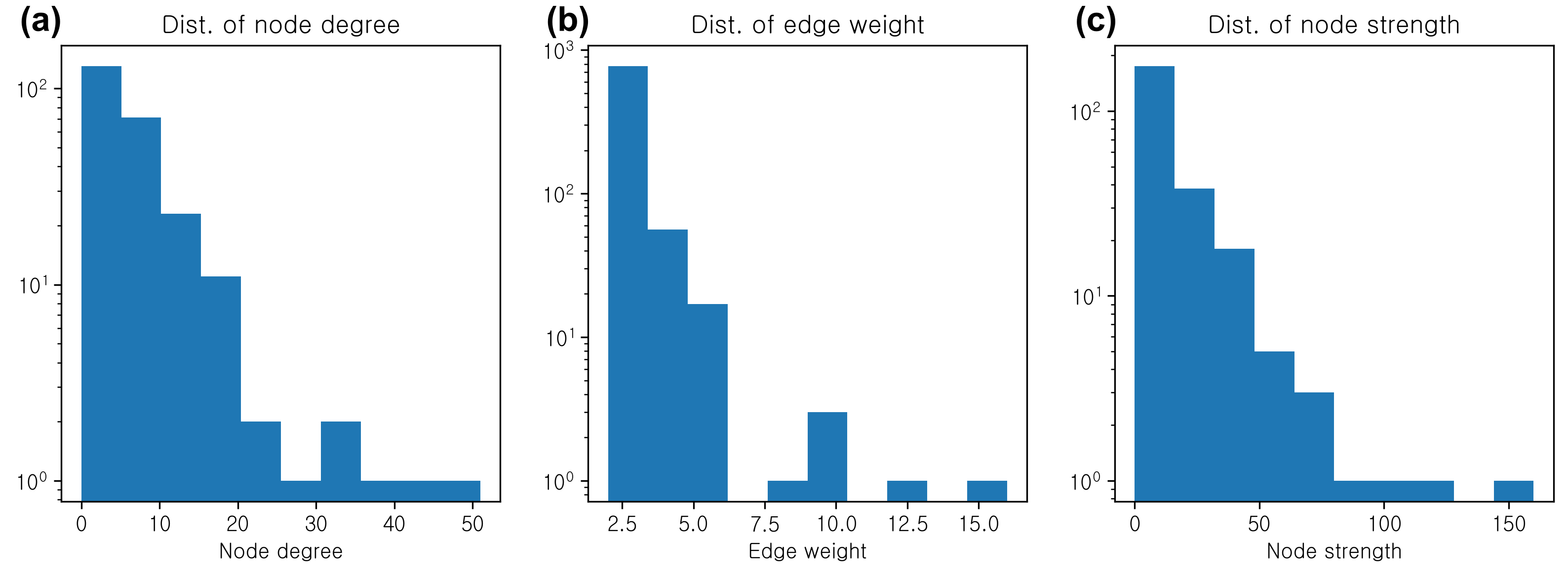}
\caption{For physics terms only, (a) degree, (b) edge-weight, and (c) node-strength distributions. Here all statistical values are obtained from the entire subjects.}
\label{fig:PhysStat}
\end{figure*}

\section{Results}
\subsection{Analysis of all words}
%

A total of 389 unique words were extracted from the 127 physics achievement standards, resulting in 3,931 word-to-word connections. As shown in Fig.~\ref{fig:TotalStat}, the distributions of node degree, node strength, and edge weight in the overall semantic network are all right-skewed. This indicates that small values occur far more frequently than large ones.
When the distribution of node degree ($k$) is analyzed using a power-law model ($p \sim k^{-\gamma}$), the exponent $\gamma$ is 2.65. This suggests that the overall semantic network exhibits a scale-free property,\footnote{It is generally considered that a network has scale-free property when $2 < \gamma < 3$.} meaning that a few hub nodes with very high connectivity exist\cite{POW}. Examples of such hubs include ‘explain,’ ‘use,’ and ‘everyday life.’
These hub words act as bridges, connecting various other words and being widely used across physics-related subjects. Therefore, they have significant meaning for physics learning.

As shown in Table~\ref{tab:Total},\footnote{All the tables are included in APPENDIX for readability.} the key words in the overall semantic network—those with high node strength and random-walk betweenness centrality—are more closely associated with scientific thinking, processes, and skills rather than specific physics concepts or terminology. In elementary school, basic skills such as ‘use’ and ‘observe’ are emphasized, whereas in middle and high school, the focus shifts towards higher-order thinking such as ‘explain,’ ‘understand,’ and ‘utilize.’
In addition, the importance of the term ‘everyday life’ diminishes with increasing grade level. This trend aligns with the differing goals of the subjects: while elementary and middle school \textbf{Science} and high school \textbf{Integrated Science} aim to foster students’ scientific literacy, \textbf{Physics}, \textbf{Mechanics and Energy}, and \textbf{Electromagnetism and Quantum} focus on developing academic literacy in physics.
Thus, the analysis results appear to be consistent with the curriculum’s intended structure. In summary, the achievement standards in the physics domain of the 2022 Revised National Curriculum reflect students’ cognitive development stages.

%

\subsection{Analysis of physics terms}

A total of 243 physics terms were identified from the achievement standards, resulting in 852 word-to-word connections. As shown in Fig.~\ref{fig:PhysStat}, the distributions of node degree, node strength, and edge weight in the physics term network are also right-skewed, similar to the overall semantic network. 
Notably, a power-law model analysis ($p \sim k^{-\gamma}$) of the node degree distribution yielded a relatively large exponent $\gamma$ of 4.08. This suggests that the physics-term network more closely resembles a random network than a scale-free network, indicating the absence of prominent hub nodes.

\begin{figure}
\begin{tabular}{ll}
(a) \\
\includegraphics[width=0.5\textwidth ]{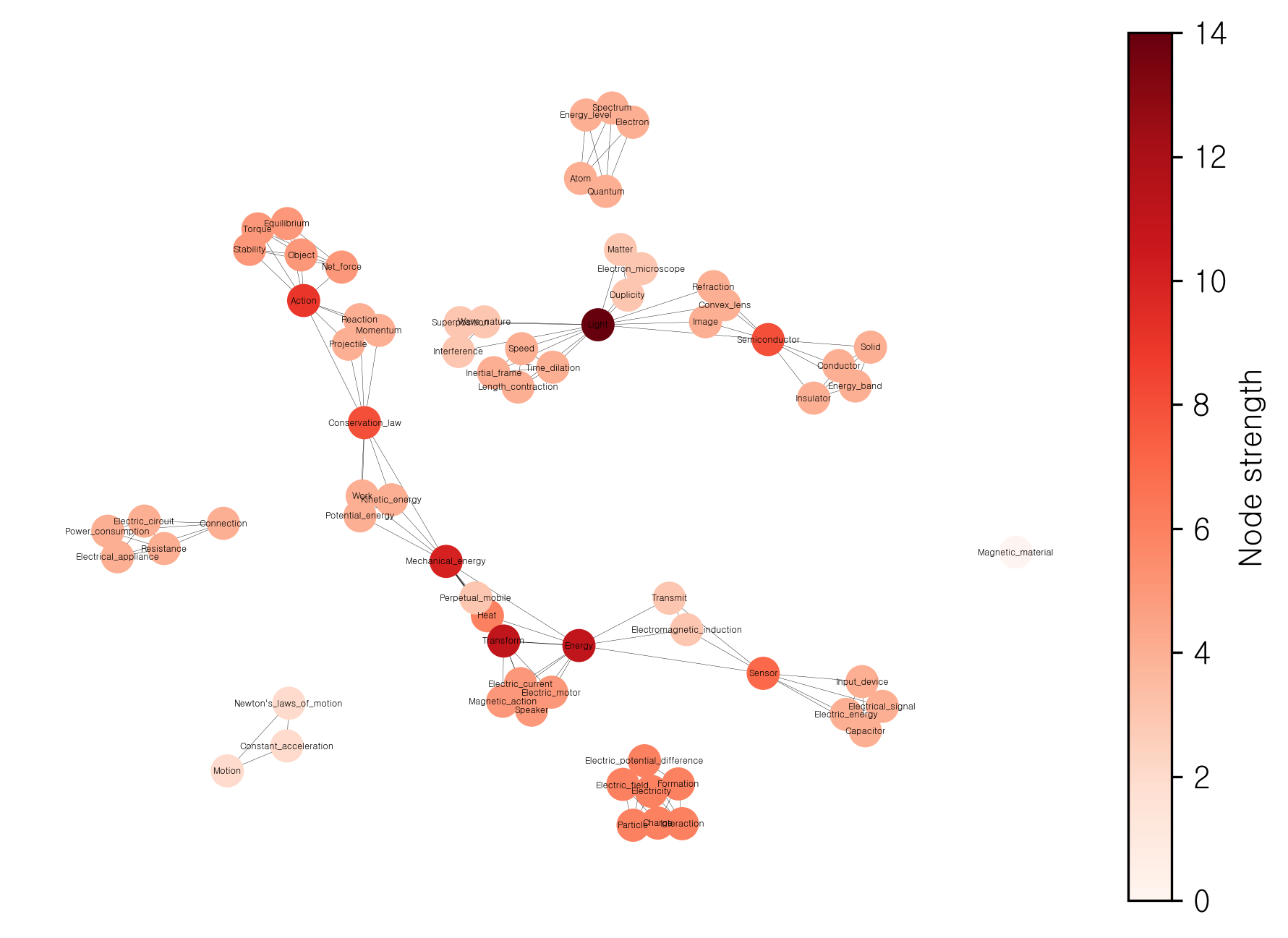}\\
(b) \\
\includegraphics[width=0.5\textwidth ]{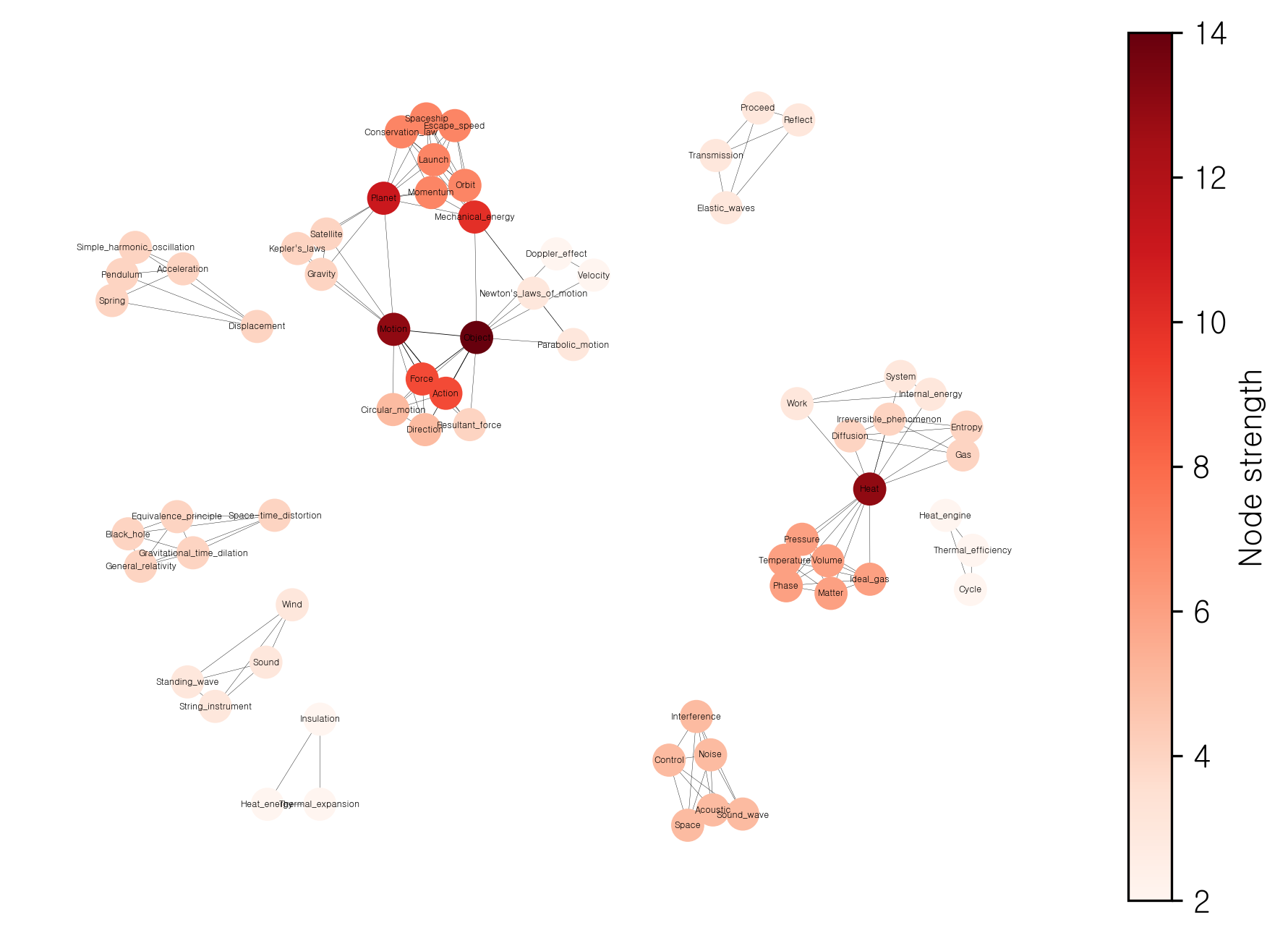}\\
(c) \\
\includegraphics[width=0.5\textwidth ]{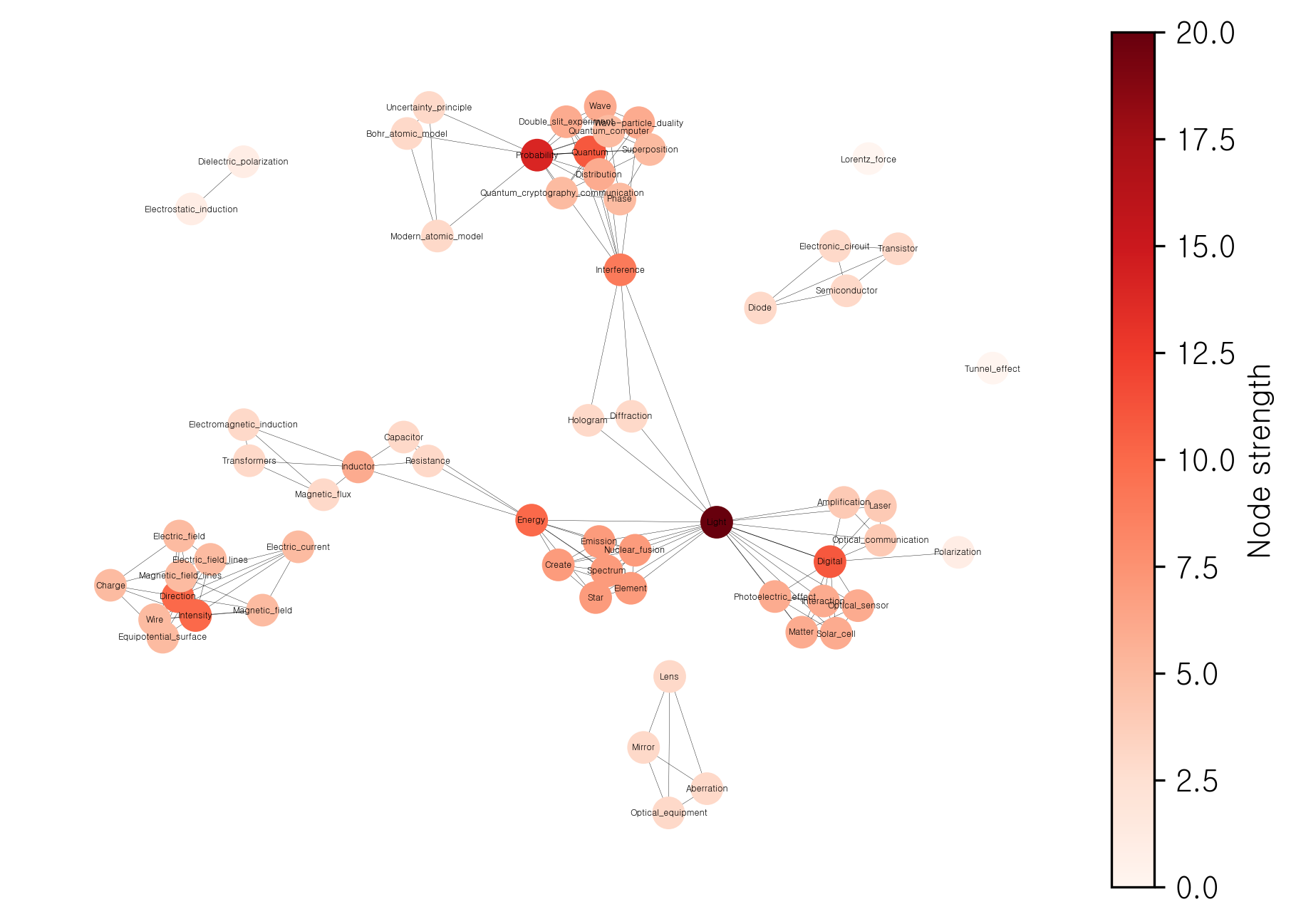}\\
\end{tabular}
\caption{Physics-term networks for three different subjects: (a) ‘Physics (물리학)’ (top), (b) ‘Mechanics and Energy (역학과 에너지)’ (middle), (c) ‘Electromagnetism and Quantum (전자기와 양자)’ (bottom). The darker the red color of a node, the stronger the node strength. The thicker the edge, the larger the edge weight.}
\label{fig:PhyWordNetworks}
\end{figure}

Before we detail the key physics terms, it is necessary to examine the structural characteristics of the network. We observed that, unlike the overall semantic network, the physics-term networks consist of disconnected components in most subjects. As shown in Fig.~\ref{fig:PhyWordNetworks}, subjects such as \textbf{Physics}, \textbf{Mechanics and Energy}, and \textbf{Electromagnetism and Quantum} are each divided into several separate components.
This fragmentation suggests that concepts within these subjects are not organically interconnected. Therefore, it implies a lack of clear conceptual coherence across physics concepts in the Korean national curriculum\cite{L_K_2022}.

\renewcommand{\thefootnote}{\fnsymbol{footnote}}
Now, let us examine the key terms in the physics-term network. Table~\ref{tab:Phys} presents the core terms for each subject. The main findings from the core term analysis can be summarized in two key points:
First, in elementary and middle school \textbf{Science} and in \textbf{Mechanics and Energy}, the core terms include classical mechanics concepts such as ‘object,’ ‘force,’ and ‘motion.’ Additionally, in \textbf{Physics}, terms such as ‘energy’ and ‘mechanical energy,’ which also belong to classical mechanics, appear as core terms. This suggests that these four subjects may be interconnected in terms of content.
Second, the key terms in \textbf{Integrated Science} and \textbf{Electromagnetism and Quantum} are distinct from those in other subjects of physics. This result indicates that these two subjects may have weaker connections with the others. In particular, \textbf{Integrated Science} primarily covers content related to astronomy, which is absent from the core terms of elementary and middle school \textbf{Science}.

\begin{figure*}
\includegraphics[width=0.75\textwidth]{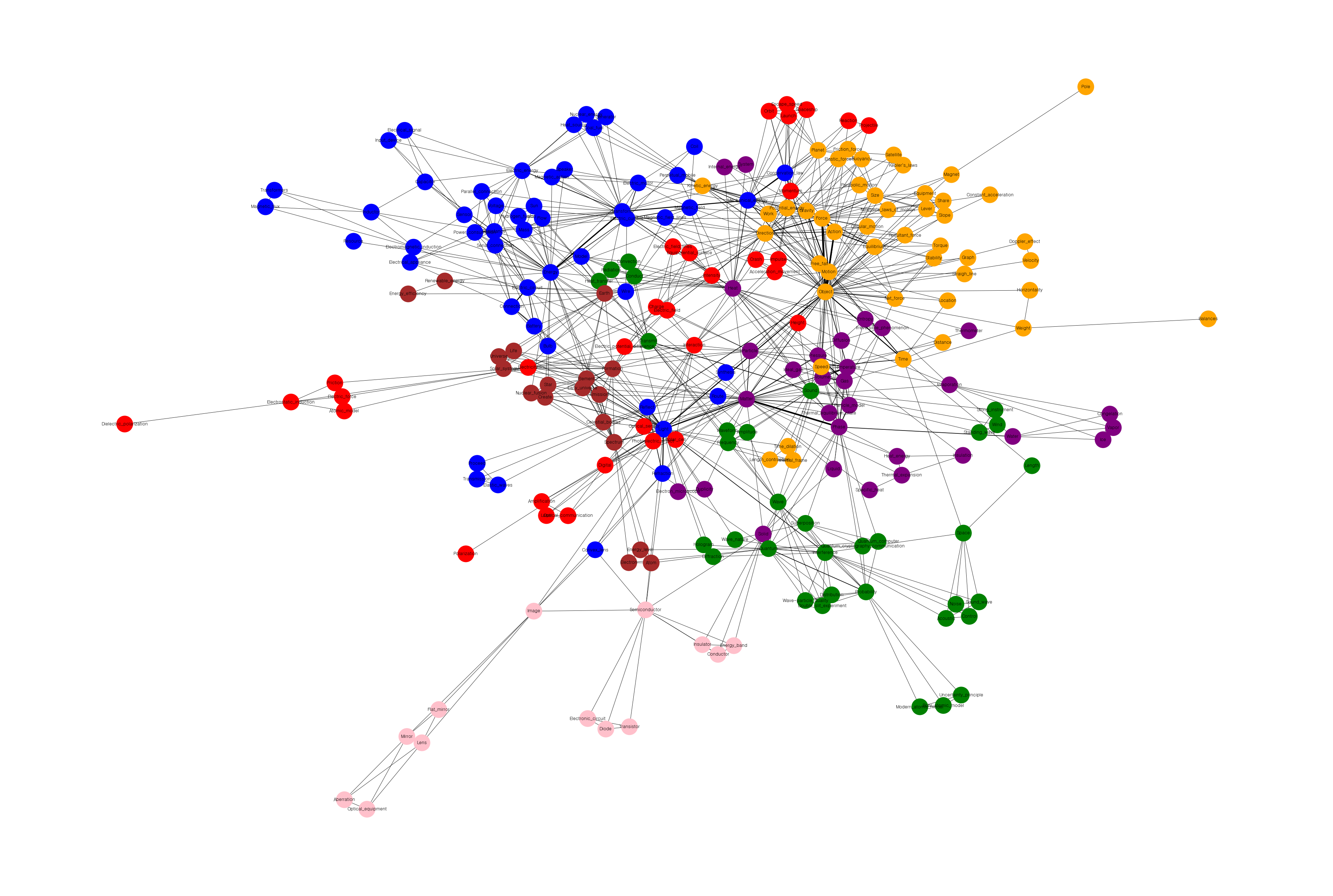}
\caption{Community structure of the physics-term network taken from the entire physics subjects. Here all nodes belong to the largest component of the physics-term network.}
\label{fig:Community}
\end{figure*}
%

\newpage
\subsection{Analysis of communities}

To explore word communities among physics terms, we conducted community detection on a semantic network constructed from all physics terms found across subjects. A total of 15 communities were identified, resulting in a modularity value of approximately 0.600. Since a modularity value greater than 0.3 is generally considered to indicate a meaningful community structure\cite{OGA}, this result suggests that the detected communities offer significant insights into the organization of physics terms.
Given the importance of understanding inter-community relationships, we focused on the seven communities belonging to the largest component of the physics term network. The words within each of these communities, along with their respective counts, are presented in Table~\ref{tab:Community}.

The results of the community detection are also visualized in Fig.~\ref{fig:Community}. As shown in the figure, related terms tend to cluster within the same communities. For example, \textbf{Community 1} includes ‘bulb,’ ‘wire,’ ‘electromagnetic induction,’ and ‘electrical signal’; \textbf{Community 2} includes ‘Newton's laws of motion,’ ‘projectile motion,’ ‘circular motion,’ and ‘potential energy’; and \textbf{Community 5} includes ‘irreversible phenomenon,’ ‘specific heat,’ ‘thermal equilibrium state,’ and ‘entropy’.
However, some physics terms are not well-grouped. For instance, ‘convex lens’ and ‘perpetual mobile’, or ‘magnet’ and ‘Doppler effect’ appear in the same community despite their conceptual disconnect. Conversely, related concepts such as ‘mechanical energy’ and ‘escape speed’ are not clustered with ‘kinetic energy’ and ‘potential energy,’ even though they are closely related.

\begin{figure*}[]
\includegraphics[width=0.8\textwidth]{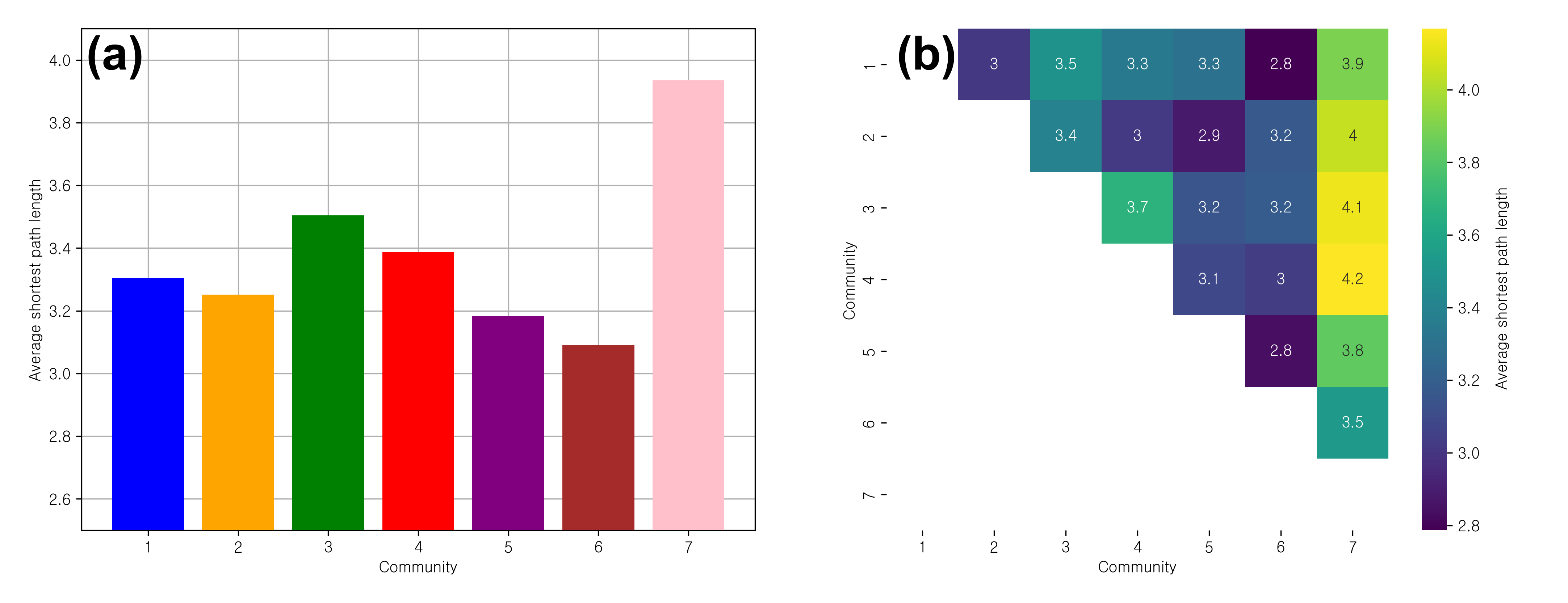}
\caption{Average shortest path length (a) per community against nodes in all other communities, and that (b) against all pairings of two communities, respectively.}
\label{fig:CommStat}
\end{figure*}
\begin{figure*}
\includegraphics[width=\textwidth]{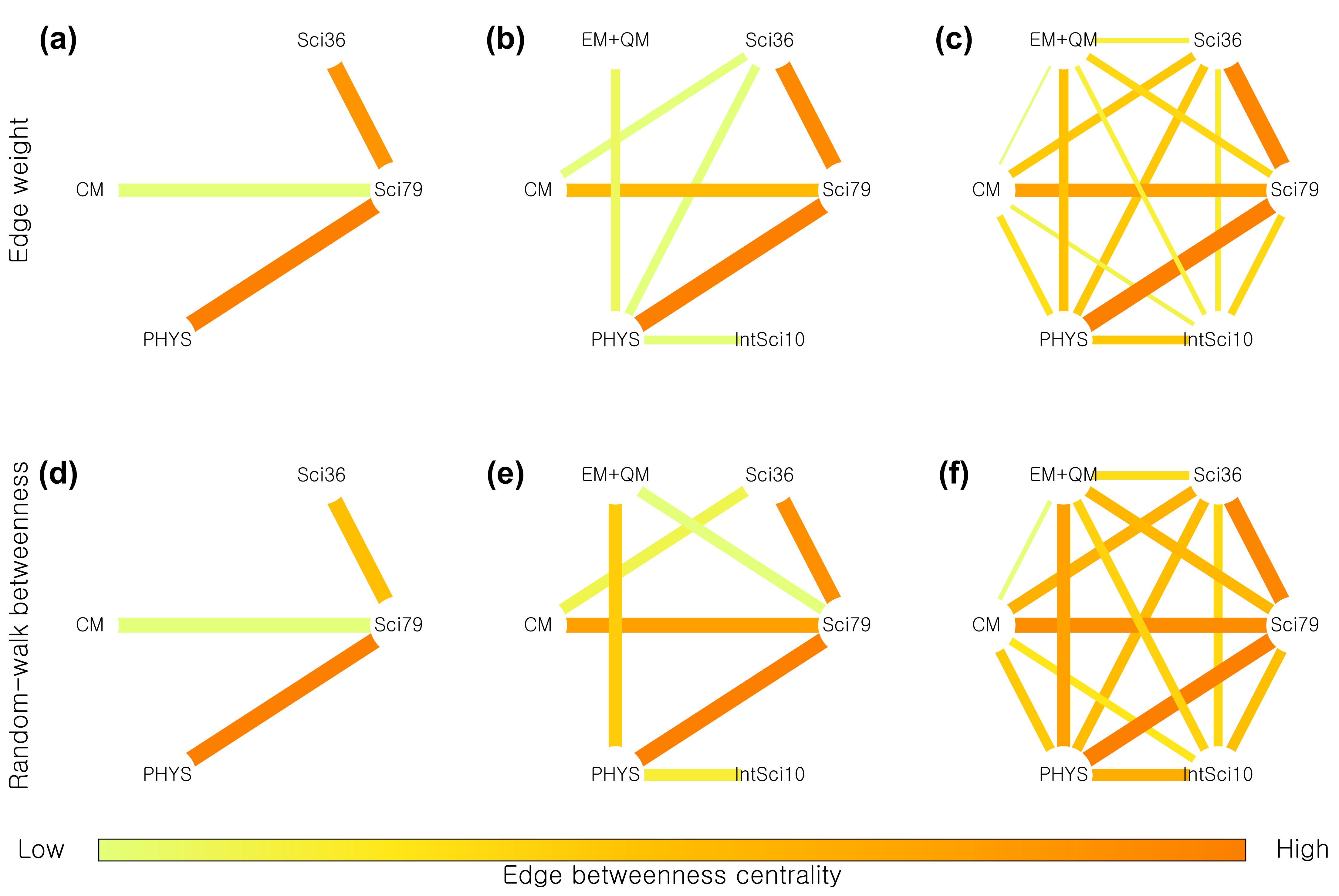}
\caption{The subject-subject correlation analyses in the subjects weighted projection: the results of edge weight [(a), (b), and (c)] and random-walk betweenness [(d), (e), and (f)], respectively. Here, the top three edges are presented for (a) and (d), the top seven edges are for (b) and (e), and all edges are for (c) and (f). The darker the red color and the thicker the line, the larger the edge betweenness centrality value: edge weight [(a)-(c)] and random-walk betweenness [(d)-(f)].}
\label{fig:Projections}
\end{figure*}

%
This result suggests that ‘mechanical energy’ is included in achievement standards related to both classical mechanics and thermodynamics, whereas ‘kinetic energy’ and ‘potential energy’ are included only in those related to classical mechanics. This probably explains why these terms are assigned to different communities.
Since simply examining the community detection may not sufficiently reveal the conceptual relationships among physics terms, we further analyzed the \textit{inter-community distances} to better understand the proximity between communities.

Figure~\ref{fig:CommStat}~(a) illustrates the average shortest path length between the nodes in a given community and randomly selected nodes from other communities. The shortest path length refers to the minimum number of steps required to traverse from one node to another in the network. As shown in Fig.~\ref{fig:CommStat}~(a), Community~6 has a relatively short average shortest path length, whereas Community~7 exhibits a longer one. This suggests that communities with shorter average shortest path lengths are more closely related to other word communities, indicating a higher level of interconnection. On the other hand, the words that make up Community~7—such as ‘mirror,’ ‘lens,’ and ‘semiconductor’—appear to have relatively weak associations with other words in the network.
Furthermore, Fig.~\ref{fig:CommStat}~(b) presents the pairwise shortest path lengths between all communities, revealing additional findings. Communities~6, 1, and 5 are closely connected, whereas Communities~7, 3, and 4 are relatively distant from each other. Communities~3, 4, and 7 include terms related to acoustics, classical mechanics, and geometrical optics, respectively. Although geometrical optics and acoustics are categorized under the same content domain in the national curriculum, the considerable distance between these communities suggests a potential lack of interconnections between them.

\subsection{Connectivity between subjects}
\label{sec:Relation}

To analyze the connectivity among subjects, we constructed a bipartite network in which the upper nodes represent physics-related subjects and the lower nodes represent physics terms. Then a projected network composed solely of subject nodes was reconstructed, and is presented as shown in Fig.~\ref{fig:Projections}. In this figure, elementary school \textbf{Science} is labeled as \texttt{Sci36}, middle school \textbf{Science} as \texttt{Sci79}, \textbf{Integrated Science} as \texttt{IntSci10}, and the high school subjects \textbf{Physics}, \textbf{Mechanics and Energy}, and \textbf{Electromagnetism and Quantum} are labeled as \texttt{PHYS}, \texttt{CM}, and \texttt{EM+QM}, respectively.
To evaluate the importance of the edges in the projected network, we used both \textit{edge weights} and the \textit{random-walk betweenness centrality of the edges}. The original projected network contained 15 edges. To identify the most significant connections, we retained only the top 3 and top 7 edges based on their edge weight and random-walk betweenness centrality values.
The findings from the analysis of this projected network can be summarized in three key points.

First, middle school \textbf{Science} serves as a core component for learning physics content in the 2022 Revised National Curriculum. As shown by the strong connections between middle school \textbf{Science} and high school subjects such as \textbf{Physics} and \textbf{Mechanics and Energy} in Fig.~\ref{fig:Projections}, much of the content taught in middle school reappears in high school physics-related subjects. Furthermore, the importance of middle school \textbf{Science} in the broader context of physics education from elementary through high school can be confirmed in Fig.~\ref{fig:Projections}~(d), which shows that middle school \textbf{Science} is connected to both elementary school \textbf{Science} and high school \textbf{Physics}.

Second, there are concerns that \textbf{Integrated Science} may not be fully fulfilling its intended curriculum role. As seen in Figures~\ref{fig:Projections}~(a), (b), (d), and (e), \textbf{Integrated Science} is either weakly connected or entirely disconnected from elementary and middle school \textbf{Science}, as well as from the high school physics-related subjects. This makes it difficult to conclude that \textbf{Integrated Science} effectively bridges the content learned in elementary and middle school with high school material.
However, this result may be attributed to the fact that \textbf{Integrated Science} has the fewest number of achievement standards. Nevertheless, the relatively low edge weights and random-walk betweenness centrality values for the edges connecting \textbf{Integrated Science} to elementary \textbf{Science} suggest that it may struggle to perform its intended bridging role.

Third, there is low connectivity among the high school subjects \textbf{Physics}, \textbf{Mechanics and Energy}, and \textbf{Electromagnetism and Quantum}. The results shown in Figures~\ref{fig:Projections}~(b) and (e) indicate that the connections between these high school physics-related subjects are either absent or very weak.

\section{Discussion}

This study constructed a semantic network based on the achievement standards of physics-related subjects in the 2022 Revised National Curriculum in South Korea. We identified the core words for each subject and analyzed word communities of physics terms to examine the interconnections among these subjects. The main findings are summarized below.

First, the keywords derived from the physics achievement standards mainly relate to scientific thinking, processes, and skills. These words demonstrate a progression toward higher-order concepts with increasing grade level. For example, in elementary school, basic inquiry skills such as ‘observe’ and ‘use’ are emphasized, whereas in high school, higher-level cognitive skills such as ‘explain’ and ‘utilize’ become more prominent. This suggests that the curriculum may have been designed with consideration of students’ cognitive developmental stages.

Second, we investigated the interconnections among physics subjects by analyzing only the physics terms extracted from the achievement standards. The analysis of key terms in the physics-term network revealed that elementary and middle school \textbf{Science}, along with high school \textbf{Physics} and \textbf{Mechanics and Energy}, share core terms from classical mechanics, suggesting that these four subjects may be closely related. In contrast, the core terms in \textbf{Integrated Science} and \textbf{Electromagnetism and Quantum} are distinct from those of the other subjects. 
Our subject connectivity analysis, using the projected network, further indicated that middle school \textbf{Science} acts as a central link, connecting elementary school \textbf{Science}, \textbf{Physics}, and \textbf{Mechanics and Energy}.
However, this analysis also revealed weak connections among the high school physics subjects themselves.
This finding aligns with prior research on the 2022 Revised Physics Curriculum\cite{I_2023}. Previous studies identified the reduction of learning content as a major issue in the 2022 revision, raising concerns about insufficient conceptual connections. Moreover, during the curriculum development, there was a debate regarding the emphasis on classical versus modern physics in the high school curriculum.
Thus, this study also confirms the possibility that efforts to reduce students’ academic burden and selectively include knowledge relevant to a rapidly changing society may have inadvertently led to a lack of conceptual coherence among high school physics subjects.

Third, our results suggest that the achievement standards for 
\textbf{Integrated Science} may be insufficient to fulfill the original purpose intended by the curriculum. \textbf{Integrated Science} is designed to connect the science content learned through middle school, foster scientific literacy, and provide foundational scientific knowledge for studying high school science and elective subjects. However, our subject connectivity analysis revealed that \textbf{Integrated Science} may not effectively bridge the content of elementary and middle school \textbf{Science} with that of high school physics-related subjects.
This outcome may stem from a disconnection in the structure and sequence of the learning content, possibly resulting from the effort to reduce the learning load by focusing on core concepts\cite{P_K_H_C_2023, L6_2024}. Therefore, it can be concluded that for \textbf{Integrated Science} to achieve the original goals of the national curriculum, collaborative efforts toward improvement by educators and researchers in the field are essential.

We anticipate that the findings of this study, which analyzed the physics domain of the 2022 Revised National Curriculum, will provide a valuable foundation for successful school education. 
We conclude by briefly discussing two limitations of this study and suggesting directions for future research to address them.

From a methodological perspective, a limitation of this study lies in our approach to identifying word communities of physics terms through network community detection.
We used an algorithm that assumes each word belongs to one community. However, this may not fully reflect reality, as a single word could belong to multiple communities.
Therefore, future research could examine the curriculum by analyzing semantic networks using overlapping communities\cite{NetSci}.

From an educational perspective, this study focused solely on the achievement standards of the curriculum, excluding analysis of other instructional media such as textbooks. However, in actual educational settings, teachers use various types of teaching media derived from the curriculum. Thus, analyzing these media is essential.
Furthermore, efforts in educational practice are needed to resolve weak connections between curriculum content through the implementation of integrated units. Integrated units aim to move away from fragmented, decontextualized instruction and instead promote convergence-based inquiry centered on themes, rather than knowledge-centered learning\cite{C_N_L_2016}. Research examining how the operation of integrated units affects student learning in the field of physics could yield meaningful insights.

\section*{Data availability}
The data, programs, and Korean-language materials used in this study are available via the following link:
\href{https://github.com/jabamseo/24Phys_curr}{https://github.com/jabamseo/24Phys$\_$curr}.

\section*{Acknowledgements}
This research was supported by the Research Fund of Chosun University (2020 On-Campus Research Grant, M.H.).

\newpage
\onecolumngrid
\section*{APPENDIX}
\begin{table*}[h]
\caption{Top ten words ranked by node strength (upper) and random-walk betweenness (lower) from all subjects to individual subjects; values are in parentheses.}
\begin{ruledtabular}
\begin{tabular}{cccccccc}
Rank & \multicolumn{1}{c}{\begin{tabular}[c]{@{}c@{}}All\\ subjects\end{tabular}} & \begin{tabular}[c]{@{}c@{}}Elementary school\\ science\end{tabular} & \begin{tabular}[c]{@{}c@{}}Middle school\\ science\end{tabular} & \begin{tabular}[c]{@{}c@{}}Integrated\\ science\end{tabular} & Physics & \begin{tabular}[c]{@{}c@{}}Mechanics\\ and energy\end{tabular} & \begin{tabular}[c]{@{}c@{}}Electromagnetism\\ and quantum\end{tabular} \\ \hline
\multirow{2}{*}{1} & Explain & Use & Explain & Various & Explain & Explain & Technology \\
 & (379) & (100) & (102) & (65) & (75) & (88) & (61) \\
\multirow{2}{*}{2} & Use & Everyday life & Everyday life & Technology & Light & Understand & Utilize \\
 & (309) & (73) & (71) & (52) & (40) & (85) & (53) \\
\multirow{2}{*}{3} & Everyday life & Inspect & Use & Earth & Utilize & Use & Light \\
 & (247) & (65) & (69) & (52) & (39) & (62) & (44) \\
\multirow{2}{*}{4} & Various & Object & Understand & Everyday life & Principle & Example & Explain \\
 & (234) & (64) & (61) & (43) & (37) & (52) & (42) \\
\multirow{2}{*}{5} & Understand & Observe & Object & Utilize & Various & Various & Principle \\
 & (222) & (64) & (61) & (42) & (35) & (49) & (38) \\
\multirow{2}{*}{6} & Utilize & Explain & Express & Measure & Apply & Utilize & Inspect \\
 & (214) & (45) & (56) & (38) & (33) & (49) & (38) \\
\multirow{2}{*}{7} & Example & Property & Feature & Nature & Example & Object & Recognize \\
 & (185) & (43) & (49) & (37) & (30) & (29) & (37) \\
\multirow{2}{*}{8} & Object & Method & Various & Meaning & Use & Inspect & Digital \\
 & (171) & (39) & (46) & (35) & (30) & (26) & (33) \\
\multirow{2}{*}{9} & Inspect & Phenomenon & Motion & Procedure & Energy & Heat & Probability \\
 & (171) & (38) & (46) & (35) & (26) & (26) & (32) \\
\multirow{2}{*}{10} & Technology & Example & Matter & Composition & Technology & Motion & Understand \\
 & (150) & (34) & (45) & (33) & (25) & (25) & (31)
\end{tabular}
\end{ruledtabular}

\begin{tabular}{l}
\end{tabular}
\begin{ruledtabular}
\begin{tabular}{cccccccc}
Rank & \multicolumn{1}{c}{\begin{tabular}[c]{@{}c@{}}All\\ subjects\end{tabular}} & \begin{tabular}[c]{@{}c@{}}Elementary school\\ science\end{tabular} & \begin{tabular}[c]{@{}c@{}}Middle school\\ science\end{tabular} & \begin{tabular}[c]{@{}c@{}}Integrated\\ science\end{tabular} & Physics & \begin{tabular}[c]{@{}c@{}}Mechanics\\ and energy\end{tabular} & \begin{tabular}[c]{@{}c@{}}Electromagnetism\\ and quantum\end{tabular} \\ \hline
\multirow{2}{*}{1} & Explain & Use & Explain & Various & Explain & Explain & Technology \\
 & (0.1571) & (0.2174) & (0.2036) & (0.1786) & (0.3552) & (0.3191) & (0.2025) \\
\multirow{2}{*}{2} & Use & Object & Everyday life & Apply & Apply & Understand & Explain \\
 & (0.1174) & (0.1860) & (0.1379) & (0.1749) & (0.1753) & (0.3165) & (0.1905) \\
\multirow{2}{*}{3} & Understand & Observe & Object & Earth & Various & Use & Use \\
 & (0.0977) & (0.1808) & (0.1314) & (0.1710) & (0.1753) & (0.1943) & (0.1648) \\
\multirow{2}{*}{4} & Everyday life & Everyday life & Use & Technology & Demonstrate & Various & Light \\
 & (0.0869) & (0.1405) & (0.1272) & (0.1391) & (0.1434) & (0.1624) & (0.1572) \\
\multirow{2}{*}{5} & Various & Inspect & Motion & Everyday life & Principle & Example & Utilize \\
 & (0.0847) & (0.1319) & (0.1245) & (0.1350) & (0.1415) & (0.1522) & (0.1540) \\
\multirow{2}{*}{6} & Utilize & Explain & Understand & Procedure & Example & Utilize & Reason \\
 & (0.0777) & (0.1238) & (0.1171) & (0.1099) & (0.1393) & (0.1484) & (0.1490) \\
\multirow{2}{*}{7} & Object & Compare & Express & Meaning & Light & Heat & Principle \\
 & (0.0708) & (0.1094) & (0.1097) & (0.1064) & (0.1211) & (0.0886) & (0.1466) \\
\multirow{2}{*}{8} & Example & Property & Matter & Utilize & Utilize & Object & Probability \\
 & (0.0631) & (0.0925) & (0.0983) & (0.0943) & (0.1070) & (0.0711) & (0.1339) \\
\multirow{2}{*}{9} & Inspect & Phenomenon & Relationship & Nature & Transform & Motion & Recognize \\
 & (0.0604) & (0.0823) & (0.0920) & (0.0905) & (0.0935) & (0.0664) & (0.1119) \\
\multirow{2}{*}{10} & Principle & Method & Procedure & Explain & Energy & Everyday life & Inspect \\
 & (0.0561) & (0.0807) & (0.0830) & (0.0854) & (0.0906) & (0.0604) & (0.0956)
\end{tabular}
\end{ruledtabular}
\label{tab:Total}
\end{table*}

\newpage
\begin{table*}[]
\caption{Top ten physics terms ranked by node strength (upper) and random-walk betweenness (lower) from all subjects to individual subjects; values are in parentheses.}
\begin{ruledtabular}
\begin{tabular}{cccccccc}
Rank & \multicolumn{1}{c}{\begin{tabular}[c]{@{}c@{}}All\\ subjects\end{tabular}} & \begin{tabular}[c]{@{}c@{}}Elementary school\\ science\end{tabular} & \begin{tabular}[c]{@{}c@{}}Middle school\\ science\end{tabular} & \begin{tabular}[c]{@{}c@{}}Integrated\\ science\end{tabular} & Physics & \begin{tabular}[c]{@{}c@{}}Mechanics\\ and energy\end{tabular} & \begin{tabular}[c]{@{}c@{}}Electromagnetism\\ and quantum\end{tabular} \\ \hline
\multirow{2}{*}{1} & Object & Object & Object & Earth & Light & Object & Light \\
 & (80) & (25) & (31) & (19) & (14) & (14) & (20) \\
\multirow{2}{*}{2} & Light & Force & Motion & Element & Energy & Motion & Probability \\
 & (56) & (10) & (27) & (13) & (11) & (13) & (14) \\
\multirow{2}{*}{3} & Motion & Matter & Force & Formation & Transform & Heat & Digital \\
 & (48) & (8) & (16) & (13) & (11) & (13) & (11) \\
\multirow{2}{*}{4} & Matter & Gas & Matter & Early universe & Mechanical energy & Planet & Quantum \\
 & (47) & (8) & (15) & (13) & (10) & (11) & (11) \\
\multirow{2}{*}{5} & Force & Light & Gravity & Transform & Act & Mechanical energy & Intensity \\
 & (35) & (8) & (12) & (12) & (9) & (10) & (10) \\
\multirow{2}{*}{6} & Energy & Battery & Transmit & Matter & Conservation law & Act & Direction \\
 & (35) & (8) & (11) & (9) & (8) & (9) & (10) \\
\multirow{2}{*}{7} & Transform & Electric circuit & Particle & Emission & Semiconductor & Force & Energy \\
 & (33) & (8) & (10) & (7) & (8) & (9) & (10) \\
\multirow{2}{*}{8} & Act & Connection & Model & Light & Sensor & Spaceship & Interference \\
 & (30) & (8) & (10) & (7) & (7) & (7) & (9) \\
\multirow{2}{*}{9} & Phase & Temperature & Particle model & Spectrum & Heat & Launch & Nuclear fusion \\
 & (26) & (7) & (10) & (7) & (6) & (7) & (7) \\
\multirow{2}{*}{10} & Heat & Lever & Direction & Celestial bodies & Interaction & Conservation law & Create \\
 & (25) & (6) & (10) & (7) & (6) & (7) & (7)
\end{tabular}
\end{ruledtabular}

\begin{tabular}{l}
\end{tabular}
\begin{ruledtabular}
\begin{tabular}{cccccccc}
Rank & \multicolumn{1}{c}{\begin{tabular}[c]{@{}c@{}}All\\ subjects\end{tabular}} & \begin{tabular}[c]{@{}c@{}}Elementary school\\ science\end{tabular} & \begin{tabular}[c]{@{}c@{}}Middle school\\ science\end{tabular} & \begin{tabular}[c]{@{}c@{}}Integrated\\ science\end{tabular} & Physics & \begin{tabular}[c]{@{}c@{}}Mechanics\\ and energy\end{tabular} & \begin{tabular}[c]{@{}c@{}}Electromagnetism\\ and quantum\end{tabular} \\ \hline
\multirow{2}{*}{1} & Light & Object & Object & Earth & Mechanical energy & Object & Light \\
 & (0.2980) & (0.7883) & (0.4567) & (0.7137) & (0.5630) & (0.4517) & (0.7113) \\
\multirow{2}{*}{2} & Object & Light & Motion & Transform & Energy & Motion & Interference \\
 & (0.2871) & (0.3135) & (0.4363) & (0.3804) & (0.4917) & (0.3770) & (0.4822) \\
\multirow{2}{*}{3} & Matter & Matter & Model & Matter & Conservation law & Planet & Energy \\
 & (0.1889) & (0.2954) & (0.3158) & (0.1313) & (0.4905) & (0.3526) & (0.3001) \\
\multirow{2}{*}{4} & Motion & Gas & Force & Formation & Act & Mechanical energy & Probability \\
 & (0.1488) & (0.2560) & (0.3098) & (0.1272) & (0.3420) & (0.3168) & (0.2685) \\
\multirow{2}{*}{5} & Energy & Phase & Transform & Early universe & Transform & Act & Inductor \\
 & (0.1466) & (0.2368) & (0.2864) & (0.1272) & (0.3385) & (0.1081) & (0.1700) \\
\multirow{2}{*}{6} & Phase & Temperature & Resistance & Element & Sensor & Force & Digital \\
 & (0.1189) & (0.2066) & (0.2645) & (0.1272) & (0.2869) & (0.1081) & (0.1687) \\
\multirow{2}{*}{7} & Transform & Water & Electric current & Energy & Heat & Kepler’s laws & Quantum \\
 & (0.1167) & (0.1669) & (0.2587) & (0.0816) & (0.1609) & (0.0843) & (0.1660) \\
\multirow{2}{*}{8} & Heat & Sound & Gravity & Hydrogen fusion & Kinetic energy & Satellite & Diffraction \\
 & (0.1131) & (0.1544) & (0.1821) & (0.0816) & (0.1127) & (0.0843) & (0.1262) \\
\multirow{2}{*}{9} & Interference & Weight & Transmit & Flow & Potential energy & Gravity & Hologram \\
 & (0.0924) & (0.1254) & (0.1646) & (0.0816) & (0.1127) & (0.0843) & (0.1262) \\
\multirow{2}{*}{10} & Transmit & Force & Matter & Mass & Work & \multicolumn{1}{c}{\begin{tabular}[c]{@{}c@{}}Newton’s laws\\ of motion\end{tabular}} & \multicolumn{1}{c}{\begin{tabular}[c]{@{}c@{}}Wave-particle\\ duality\end{tabular}} \\
 & (0.0858) & (0.1145) & (0.1579) & (0.0816) & (0.1127) & (0.0829) & (0.0697)
\end{tabular}
\end{ruledtabular}
\label{tab:Phys}
\end{table*}

\newpage
\begin{table*}[]
\caption{Community detection results for the physics-term networks taken from the entire physics subjects. Here all terms belong to the largest component of the physics-term networks.}
\resizebox{\textwidth}{!}{%
\begin{ruledtabular}
\begin{tabular}{ccp{15cm}}
Community & Size & \multicolumn{1}{c}{Words} \\
\hline
1 & 51 & Battery, Speaker, Parallel connection, Transformers, Electric current, Reflect, Resistance, Synthesis, Electric circuit, Light, Mass, Coal fuel, Resource, Transform, Power consumption, Input device, Conservation law, Inductor, Flow, Electric motor, Electrical signal, Magnetic flux, Magnetic field lines, Connection, Elastic waves, Route, Transmission, Perpetual mobile, Energy, Electric energy, Nuclear energy, Proceed, Generator, Hydrogen fusion, Magnetic action, Refraction, Electromagnetic induction, Capacitor, Model, Convex lens, Electrical appliance, Magnetic field, Serial connection, Wire, Sun, Sensor, Coil, Bulb, Voltage, Heat source, Mechanical energy \\
\hline
2 & 46 & Motion, Friction force, Resultant force, Time, Object, Circular motion, Potential energy, Equipment, Horizontality, Kinetic energy, Location, Elastic force, Net force, Balances, Inertial frame, Share, Free fall, Force, Planet, Length contraction, Slope, Pole, Speed, Graph, Projectile motion, Constant acceleration, Doppler effect, Straight line, Act, Magnet, Satellite, Work, Time dilation, Weight, Size, Velocity, Lever, Newton’s laws of motion, Kepler’s laws, Gravity, Distance, Torque, Equilibrium, Stability, Direction, Buoyancy \\
\hline
3 & 34 & Uncertainty principle, Double slit experiment, Interference, Bohr atomic model, Control, Wave nature, String instrument, Noise, Transmit, Probability, Heat transfer, Space, Quantum cryptography communication, Quantum computer, Sound, Standing wave, Wave, Wind, Quantum, Diffraction, Frequency, Amplitude, Conduct, Sound wave, Hologram, Wave-particle duality, Superposition, Waveform, Length, Acoustic, Distribution, Radiation, Convection, Modern atomic model\\
\hline
4 & 32 & Solar cell, Height, Atomic model, Amplification, Spaceship, Electric potential difference, Polarization, Momentum, Digital, Interaction, Crash, Laser, Electrostatic induction, Electric field, Charge, Launch, Photoelectric effect, Intensity, Orbit, Electricity, Electric force, Friction, Equipotential surface, Acceleration movement, Dielectric polarization, Reaction, Optical communication, Projectile, Impulse, Electric field lines, Escape speed, Optical sensor \\
\hline
5 & 30 & Electron microscope, Particle model, Insulation, Specific heat, Internal energy, Duplicity, Solid, Diffusion, Matter, Phase, Heat energy, Evaporation, Vapor, Entropy, Particle, Water, Volume, Ice, Ideal gas, Pressure, System, Thermometer, Liquid, Heat, Irreversible phenomenon, Thermal equilibrium state, Temperature, Gas, Congelation, Thermal expansion \\
\hline
6 & 18 & Nuclear fusion, Emission, Solar system, Element, Star, Renewable energy, Early universe, Earth, Electron, Create, Atom, Life, Energy efficiency, Formation, Universe, Energy level, Spectrum, Celestial bodies \\
\hline
7 & 13 & Energy band, Lens, Insulator, Conductor, Optical equipment, Image, Electronic circuit, Aberration, Diode, Transistor, Flat mirror, Mirror, Semiconductor
\end{tabular}%
\end{ruledtabular}
}
\label{tab:Community}
\end{table*}



\begin{thebibliography}{10}

\bibitem{2022} Ministry of Education, The Science Curriculum, Statute Notice of Ministry of Education No. 2022-33, Sejong: Ministry of Education (2022).

\bibitem{2015} Y. Kwak, Monitoring Study on the Implementation of the 2015
National Science Curriculum in Elementary and Secondary
Schools, 2019.

\bibitem{L_K_2019} B. Lee and K. Kang, Analysis on Achievement Standards of the 2015 Revised Curriculum and Learning Objectives of Physics I Textbooks Based on Bloom’s Revised Taxonomy,
    \refdoi{New Phys.: Sae Mulli \textbf{73}, 941 (2023)}{10.3938/NPSM.73.941} 

\bibitem{K_2023} K. Kang, Comparison of Achievement Standards in Physic of 2015 and 2022 Revised Curriculum - Based on Bloom’s Revised Taxonomy of Educational Objectives,
    \refdoi{New Phys.: Sae Mulli \textbf{73}, 941 (2023)}{10.3938/NPSM.73.941} 

\bibitem{J_2023} H. Jho, Comparing the 2015 with the 2022 Revised Primary Science Curriculum Based on Network Analysis,
    \refdoi{JKESE \textbf{42}, 178 (2023)}{10.15267/keses.2023.42.1.178} 

\bibitem{J_N_C_2023} H. Jho, H. Noh and J. Choi, Network Analysis of Achievement Standards Relevant to the Concept of Energy Presented in the 2022 Revised National Curriculum of Science,
    \refdoi{JECCE \textbf{13}, 125 (2023)}{10.22368/ksecce.2023.13.2.125} 

\bibitem{Y_P_2014} E. Yun and Y. Park, Articulation Analysis of Electromagnetism Units in Middle-school Science and Physics I and II Courses in the 2009 National Curriculum,
    \refdoi{New Phys.: Sae Mulli \textbf{64}, 458 (2014)}{10.3938/NPSM.64.458}     

\bibitem{L_L_2015} J. Lim and B. Lee, Analysis of High-school Students’ Difficulty Related to Conceptual Knowledge in Solving Problems in Classical Mechanics,
    \refdoi{New Phys.: Sae Mulli \textbf{65}, 333 (2015)}{10.3938/NPSM.65.333} 

\bibitem{R_N_I_2018} S. Rhee, I. Nam and S. Im, A Relationship between Scientific Key Competencies and Achievement Standards in Physics under the 2015 Revised National Curriculum of Korea,
    \refdoi{New Phys.: Sae Mulli \textbf{68}, 1081 (2018)}{10.3938/NPSM.68.1081}

\bibitem{S_2018} Y. Song, Semantics Network Analysis of the Activation Words for ‘Energy’ Terminology Used by Secondary Pre-Service Science Teachers,
    \refdoi{New Phys.: Sae Mulli \textbf{68}, 889 (2018)}{10.3938/NPSM.68.889}

\bibitem{S_K_2020} K. Seo and Y. Kim, Changes in Articulation of Stimulus and Reaction Unit Concepts according to Revision of Curriculum Using Semantic Network Analysis,
    \refdoi{BE \textbf{48}, 20 (2020)}{10.15717/bioedu.2020.48.1.20}

\bibitem{I_2023} S. Im, 2022 Revised Curriculum and Secondary School Physics Education,
    \refdoi{Physics and High Technology \textbf{} (2023)}{10.3938/PhiT.32.025}

\bibitem{khaiii} khaiii(Kakao Hangul Analyzer III),
  \url{https://github.com/kakao/khaiii}

\bibitem{phyWord} 물리학용어집(Physics Glossary),
  \url{https://edss.moe.go.kr/man/MainRM.do}

\bibitem{RWB} M. E. J. Newman, A measure of betweenness centrality based on random walks,
    \refdoi{Soc. Netw. \textbf{27}, 39 (2005)}{10.1016/j.socnet.2004.11.009}

\bibitem{OGA} A. Clauset, M. E. J. Newman and C. Moore, Finding community structure in very large networks,
    \refdoi{PRE \textbf{70}, 066111 (2004)}{10.1103/PhysRevE.70.066111}

\bibitem{1st} F. Menczer, S. Fortunato and C. A. Davis, 
    \textit{A first course in network science} (Cambridge University Press, Cambridge, 2020), pp. 163-172.

\bibitem{NetworkX} A. Hagberg, P. Swart and D. S Chult, Exploring network structure, dynamics, and function using NetworkX. No. LA-UR-08-05495; LA-UR-08-5495. Los Alamos National Lab.(LANL), Los Alamos, NM (United States), 2008.

\bibitem{ERWB} M. Girvan and M. E. J. Newman, Community structure in social and biological networks,
    \refdoi{PNAS \textbf{99}, 7821 (2002)}{10.1073/pnas.122653799}

\bibitem{POW} Barabási, A.L. and R. Albert, Emergence of scaling in random networks,
    \refdoi{Science \textbf{286}, 509 (1999)}{10.1126/science.286.5439.509}

\bibitem{L_K_2022} J. Lee and J. B. Kim, A Comparative Study on the Concept of Light Presented in Elementary School Science Curriculum and Textbooks in Korea, the US, China, and Japan,
    \refdoi{JKESE \textbf{41}, 283 (2022)}{10.15267/keses.2022.41.2.283}

\bibitem{P_K_H_C_2023} D. Park, M. Kwon, S. Ha and H. M. Choi, A Study on the Vertical Articulation of Physics in the 2015 Revised National Curriculum: Focused on the Middle School Science and the High School Integrated Science,
    \refdoi{New Phys.: Sae Mulli \textbf{73}, 515 (2023)}{10.3938/NPSM.73.515}

\bibitem{L6_2024} B. Lee, J. Park, J. Son, K. Y. Lee, W. Choi \textit{et al.}, Concerns and Difficulties in Applying the National Curriculum in the Process of Developing Science Textbooks: Focused on ‘Integrated Science’ of the 2022 Revised National Science Curriculum,
    \refdoi{JKASE \textbf{44}, 219 (2024)}{10.14697/jkase.2024.44.2.219} 

\bibitem{NetSci} A. L. Barabási, 
    \textit{Network science} (Cambridge University Press, Cambridge, 2016), pp. 346-348.

\bibitem{C_N_L_2016} H. Chae, S. Noh and S. Lee, Analysis of Characteristics of Material–Centered Integrated Unit in Finland Elementary Science Textbook,
    \refdoi{JKESE \textbf{35}, 26 (2016)}{10.15267/keses.2016.35.1.026}
\end{thebibliography}
\end{document}